# Enhanced Magnetic Resonance Image Synthesis with Contrast-Aware Generative Adversarial Networks

Jonas Denck[a, b, c], Jens Guehring[b], Andreas Maier[a] and Eva Rothgang[c]

[a] Pattern Recognition Lab, Department of Computer Science, Friedrich-Alexander Universität Erlangen-Nürnberg, Erlangen, Germany

[b] Department of Industrial Engineering and Health, Technical University of Applied Sciences Amberg-Weiden, Weiden, Germany

[c] Siemens Healthcare GmbH, Erlangen, Germany

*Abstract*—A Magnetic Resonance Imaging (MRI) exam typically consists of the acquisition of multiple MR pulse sequences, which are required for a reliable diagnosis. With the rise of generative deep learning models, approaches for the synthesis of MR images are developed to either synthesize additional MR contrasts, generate synthetic data, or augment existing data for AI training. While current generative approaches allow only the synthesis of specific sets of MR contrasts, we developed a method to generate synthetic MR images with adjustable image contrast. Therefore, we trained a generative adversarial network (GAN) to generate synthetic MR knee images conditioned on various acquisition parameters (repetition time, echo time, image orientation). In a visual Turing test, two experts mislabeled 40.5% of real and synthetic MR images, demonstrating that the image quality of the generated synthetic and real MR images is comparable. This work can support radiologists and technologists during the parameterization of MR sequences by previewing the yielded MR contrast, can serve as a valuable tool for radiology training, and can be used for customized data generation to support AI training.

## 1. Introduction

In Magnetic Resonance Imaging (MRI), multiple contrasts are usually acquired within a single exam that are required to make a reliable diagnosis. Each contrast is based on the parameterization of an MR sequence through multiple acquisition (pulse sequence) parameters. In general, the acquisition parameters for an MR sequence affect image contrast, image resolution, signal-to-noise ratio, and/or scan time. Important acquisition parameters that affect the image contrast are the repetition time (TR) and echo time (TE). The parameterization of a sequence is subject to clinical guidelines, the MR system (vendor, model, software version, field strength), the clinical protocol (i.e., the set of parameterized sequences used for the exam), internal guidelines (e.g., slot time), and radiologists' preferences. Clinical guidelines (e.g., ACR–SPR–SSR Practice Parameters [1]) offer recommendations but "are not inflexible rules or requirements of practice and are not intended, nor should they be used, to establish a legal standard of care" [1]. Thus, the guidelines do not specify exact acquisition parameter settings as they will depend on the field strength and desired contrast weighting.

Consequently, sequence parameterizations differ significantly across different radiology sites [2, 3]. However, due to the various degrees of freedom in protocol configuration and sequence parameterization, this can also be true for a single radiology site. These differences additionally increase the complexity of MRI protocoling (i.e., selecting a set of parameterized sequences for an MRI exam), image interpretation/diagnostics, and the development of AI (artificial intelligence) applications for MRI. Since most AI applications are trained and evaluated on a limited set of MR sequences with fixed or narrowly defined acquisition parameter values [4], the applicability of AI-based applications to sequences with different parameterizations is not guaranteed. Consequently, re-training with a new set of MR images may be necessary. However, since the abundance of clinical images for AI training is limited, this is not always possible.

To mitigate the problem of the availability of MR images for varying contrast settings, we developed an approach for MR image synthesis that can be parameterized with acquisition parameters. Our method can generate customized training data for AI applications and serve as an additional data augmentation tool through contrast augmentation. Moreover, it can be used as a tool for MRI training for technologists and radiologists and can support protocoling and sequence parameterization by visualizing the yielded contrast of a parameterized sequence.

We developed a generative adversarial network (GAN) that can generate synthetic MR images of the knee conditioned on the repetition and echo time. The model can additionally synthesize MR images conditioned on the image orientation. In contrast to existing AI-based approaches for the synthesis of MR images, our model is not conditioned on different sets of contrasts (e.g., T1w, T2w, PDw) but on the acquisition parameters that determine the contrast weighting. It allows the generation of MR images with fine-tuned contrast, adjusted to the application's specific needs.

## 2. Background

Generative adversarial networks [5] learn to generate images through the adversarial training of a generator network $G$, which is trained to produce realistic samples and tries to fool the discriminator network $D$ that learns to distinguish between real and synthetic samples. Commonly, the generator takes as input a latent/noise vector $z$, randomly sampled from a normal distribution. The original GAN formulation has been adapted in several ways, e.g., introducing conditional GANs [6, 7], adapting the input for image-to-image [8], and text-to-image [9] translations, enhancing the network architecture [10] the loss formulation [11, 12], and training procedure [13]. GANs are successfully applied to various domains, and a detailed overview of GANs and their variants is given in [14].

GANs have also been successfully applied to the field of medical imaging, with applications that can mainly be divided into seven categories: synthesis, segmentation, reconstruction, detection, denoising, registration, and classification, whereas the majority of publications address synthesis applications [15].

In the following, we review publications focused on synthesizing MR images with noise-to-image GANs (Table 1). A detailed general overview of GANs in the field of medical imaging is given in [16].

Most publications in the field of MR image synthesis with noise-to-image GANs address the generation of different contrasts (e.g., T1w, T2w, PDw) in a categorical manner, without the incorporation of the acquisition parameter values. Their main target application is enhanced data augmentation for deep learning applications (see Table 1).

In [17], a Laplacian GAN [18] was trained on 2D image slices from sagittal brain MR 3D-T1w slices to enhance the augmentation of biomedical datasets.

A semi-coupled GAN was used as a data generation method for deep learning-based detection of incomplete left-ventricle coverage [19]. Additionally, a Wasserstein-GAN was used to synthesize brain MR images of different contrasts (T1w, T1c, T2w, FLAIR) with a resolution of 128×128 pixels [20] and a progressive growing GAN (PGGAN) [13] to both synthesize brain MR images and place brain metastases on synthetic MR images (256×256 pixels) for enhanced data augmentation for AI training [21, 22]. Multi-modal MR images were synthesized with a PGGAN in [23]. Moreover, GANs were used for enhanced image denoising [24] for brain MR images and the generation of additional training data for a brain tissue segmentation network [25].

**Table 1**
Literature review: MR Image Synthesis using noise-to-image GANs

| Ref. | Anatomy | Method/Network Architecture | Sequence Specification | Resolution [pixels] | Application |
|------|---------|-----------------------------|------------------------|---------------------|-------------|
| [17] | Brain | LAPGAN | T1w (4/2000ms) | 128*64 | DA |
| [26] | Heart | SCGAN | Cine | 120*120 | DA |
| [20] | Brain | DCGAN/WGAN | T1w, T1c, T2w, FLAIR (BRATS 2016 [27]) | 128*128 | DA |
| [21] | Brain | CPGGAN | T1c (BRATS 2016) | 256*256 | DA |
| [22] | Brain | PGGAN+MUNIT/SimGAN | T1c (BRATS 2016) | 256*256 | DA |
| [23] | Brain | PGGAN | T1w, T1c, T2w, FLAIR (BRATS 2016) | 256*256 | DA, unsupervised classification of pathology |
| [24] | Brain | DCGAN | T1w | 220*172 | Image denoising |
| [28] | Brain | PGGAN | FLAIR | 128*128 | Segmentation |

Legend:
DA: Data augmentation
FLAIR: Fluid-attenuated inversion recovery sequence [29]
T1c: T1-weighted contrast-enhanced sequence
LAPGAN: Laplacian GAN
SCGAN: Semi-coupled GAN
DCGAN: Deep convolutional GAN [10]
WGAN: Wasserstein GAN [12]

PGGAN: Progressive growing GAN
CPGGAN: Conditional progressive growing GAN
MUNIT: Multimodal unsupervised image-to-image translation framework [30]
SimGAN: Semantic image manipulation using generative adversarial networks [31]

Performance benefits through additional GAN-based data augmentation are reported for different medical deep learning applications [32, 33]. However, the improvement gain depends on the amount of real data seen during training, with the greatest improvements observed for training procedures with a limited amount of real data [28].

Current GAN-based MR image and contrast synthesis methods only generate single categorical contrasts with no scale of similarity between the different contrasts [17, 21, 22, 28, 30]. Consequently, the GAN does not learn to disentangle the underlying anatomy from the contrast specifications, limiting the GAN's capability for MR image synthesis. We solve this by conditioning the GAN on the acquisition parameters that determine the MR image contrast and therefore disentangling anatomy and contrast synthesis.

## 3. Material and Methods

### 3.1 Generative Adversarial Network

#### 3.1.1 Progressive Growing GAN

In order to train a network on the synthesis of MR images, we use a progressive growing GAN proposed by [13] that starts with learning low-resolution images and progressively increases the resolution by stacking additional layers to the network. Progressive growing allows the network to learn large-scale features of the data distribution first and refined structures when adding additional layers and progressing in training, stabilizing GAN training significantly. Our network architecture is trained with 800k images before doubling the resolution (following the training procedure proposed in [13]). A new layer is faded in during training with additional 800k images until we reach the final resolution of 256×256. We use a Wasserstein GAN with gradient penalty loss (WGAN-GP) [12] following the proposed training procedure in [13]:

$$L_{WGAN-GP} = \mathbb{E}_{\tilde{x} \sim P_g}[D(\tilde{x})] - \mathbb{E}_{x \sim P_r}[D(x)] + \lambda_{gp} \mathbb{E}_{\hat{x} \sim P_{\hat{x}}}[(\|\nabla_{\hat{x}} D(\hat{x})\|_2 - 1)^2] \qquad (1)$$

where $\mathbb{E}[\cdot]$ denotes the expected value, $P_r$ is the real data distribution, $P_g$ is the data distribution generated through $\tilde{x} = G(z, c)$, $c$ denotes the target labels (conditions). A gradient penalty term, weighted by $\lambda_{gp}$ ($\lambda_{gp} = 10$), is added for the random sample $\hat{x} \sim P_{\hat{x}}$, with $\nabla_{\hat{y}}$ denoting the gradient operator towards the generated samples. $P_{\hat{x}}$ describes the distribution of points uniformly sampled along straight lines from pairs of points from $P_r$ and $P_g$ [11]. The utilization of a gradient penalty term results in a more stable training process than weight clipping of the discriminator as proposed in [12].

#### 3.1.2 Separate Auxiliary Classifier

We deviate from the conventional Auxiliary Classifier GAN (ACGAN) network architecture [7], which uses a classification layer in the discriminator to learn the conditions by employing a separate auxiliary classifier (AC) that is only trained on the conditions. This allows us to use data augmentation on the training data for the AC. In general, this enhances the classification performance of a trained network by avoiding overfitting [34, 35], and only a well-trained auxiliary classifier can provide good guidance for the training of the generator. In contrast to training the AC, heavy data augmentation diminishes the GAN performance of generating sharp, realistic images. Consequently, no data augmentation is used for training the discriminator that learns to score the realness of a given image in a WGAN-GP.

We use the Xception [36] architecture for the AC and jointly train the network to determine TR, TE, and imaging orientation of the patient (IOP) from MR images only. The categorical cross-entropy loss ($CCE$) is used for the image orientation and the mean squared error ($MSE$) loss for TR and TE, which are both scaled to values between 0 and 1. The AC is trained to minimize the following loss:

$$L_{AC} = \lambda_{IOP} CCE_{IOP}(\mathbf{c}, \tilde{\mathbf{c}}) + \lambda_{TE} MSE_{TE}(\mathbf{c}, \tilde{\mathbf{c}}) + \lambda_{TR} MSE_{TR}(\mathbf{c}, \tilde{\mathbf{c}}) \tag{2}$$

where $\tilde{\mathbf{c}} = \text{C}(\mathbf{x})$ is the output of the auxiliary classifier for an image $\mathbf{x}$ with labels $\mathbf{c}$. We heuristically set $\lambda_{IOP} = 1$ and $\lambda_{TE} = \lambda_{TR} = 10$.

### 3.1.3 Controllable GAN

To avoid overfitting the generator to the conditioning, we incorporate the training procedure of ControlGAN [37]. Overfitting can occur if the auxiliary classifier reaches imperfect classification performance and consequently an issue for many machine learning-based classification and regression tasks. We use an adaptive loss weighting to avoid overfitting the generator on the conditions and balance the GAN training to produce realistic images. An adaptive weight loss parameter $\gamma_t$ for time step $t$ for each condition $c$ of the GAN is introduced:

$$\gamma_{c,t} = \min\left[\tau_{c,\gamma}, \max\left[0, \gamma_{c,t-1} + \text{r} \cdot \left\{L_c\left(\mathbf{c}, \text{C}(G(\mathbf{z}, \mathbf{c}))\right) - \widehat{\text{E}} \cdot L_c\left(\mathbf{c}, \text{C}(\mathbf{x})\right)\right\}\right]\right] \tag{3}$$

where $r$ is a learning rate parameter for $\gamma_t$, $\gamma_0$ is set to zero, and $\tau_\gamma$ is a maximum constraint for $\gamma_t$. The parameter $\tau_\gamma$ is set to 100 and the learning rate $r$ for learning of $\gamma_t$ to 0.01 for our training procedure. $\widehat{\text{E}}$ balances the ratio between the classification loss on the real training images and the generated images and is set to one. $\text{L}_c$ denotes the condition-specific loss ($CCE$ for the image orientation or $MSE$ for TR and TE). Thus, the GAN is trained to minimize the AC loss:

$$L_{AC-GAN} = \gamma_{IOP,t} CCE_{IOP}\big(\mathbf{c}, \text{C}(G(\mathbf{z}, \mathbf{c}))\big) + \gamma_{TE,t} MSE_{TE}\big(\mathbf{c}, \text{C}(G(\mathbf{z}, \mathbf{c}))\big) \tag{4}$$
$$+ \gamma_{TR,t} MSE_{TR}\big(\mathbf{c}, \text{C}(G(\mathbf{z}, \mathbf{c}))\big)$$

### 3.1.4 Overall Loss Function

Minimizing the sum of the WGAN-GP loss (1) and the auxiliary classifier loss with adaptive weights (4) leads to the generation of realistic MR images with the intended MR contrast. Thus, the overall GAN loss function is given as:

$$L_{GAN} = L_{WGAN-GP} + L_{AC-GAN} \tag{5}$$

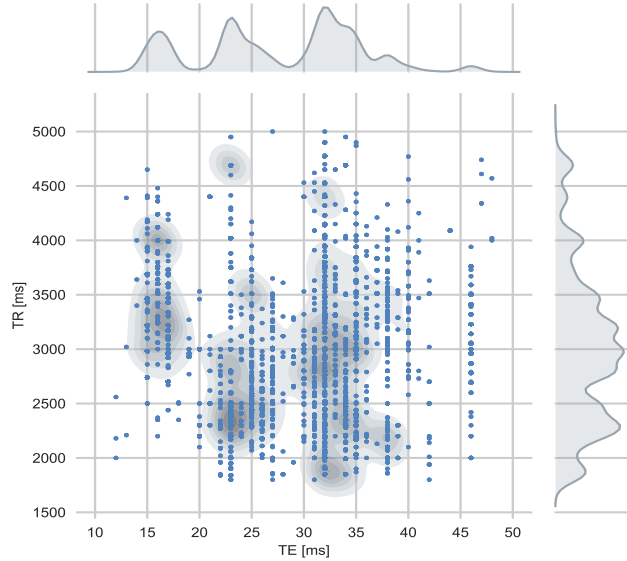

**Fig. 1** Distribution of the acquisition parameters TR and TE in the training dataset. The kernel density estimate plot visualizes the density of the bivariate target distribution. The multiple modes of the multimodal distribution (compare distribution of TE values), arise from varying sequence parameterizations used at different scanners in the dataset.

## 4. Results

### 4.1 Data

For training and evaluation of our GAN, we use the fastMRI dataset [38]. It contains DICOM data from 10,000 clinical knee MRI studies, each comprising a set of multiple pulse sequence parameterizations. We apply several data filters based on the DICOM header information to get a dataset with a comparable image impression, a dense and homogenous acquisition parameter distribution (TR, TE), and high variance in anatomy. We want the image impression and contrast within our training set to depend on the acquisition parameters TR and TE. Therefore, other parameters affecting the image impression, such as field strength and manufacturer, are removed by selecting the most common parameter value within the fastMRI dataset (1.5T field strength and scanners from Siemens Healthcare, Erlangen, Germany). Several attributes were missing in the DICOM header, but we could deduce the manufacturer for DICOM images with the missing manufacturer by adopting the manufacturer of the used receiver coil. The MR images from our filtered dataset were acquired on five different Siemens Healthcare scanners (MAGNETOM Aera, MAGNETOM Avanto, MAGNETOM Espree, MAGNETOM Sonata, MAGNETOM Symphony). To create a dense data distribution for the conditioning parameters TR and TE, we take only image series with TR values between 1800 ms and 5000 ms, set the upper limit of TE to 50 ms, and discarded non-fat saturated images. Then, we take the six central slices from each volume to discard peripheral slices. The final dataset contains MR images from 5,387 different studies and 8,535 image series. The joint distribution of TR and TE values in the training dataset is shown in Fig. 1.

The dataset of 51,205 images is split into a training, validation, and test dataset randomly by study IDs -

with 2,000 images each used for validation and testing. The remaining images were used for training. Each slice is normalized to intensity values between -1 and 1 and resized to the smallest common resolution within the dataset (256×256 pixels) using bilinear interpolation to obtain identical image resolution. Thus, the complete dataset can be used for training without discarding images due to insufficient resolution and image upsampling is avoided, which reduces image quality.

## 4.2 Training Details

We train the discriminator and generator networks adversarially with a balanced number of weight updates (Fig. 2). No conditioning is applied during the progressive growing of the networks until the final resolution is reached. The AC loss is incorporated with adaptive loss weights as defined in (3). The AC is pre-trained on the same dataset as the GAN for the final resolution for 200 epochs (batch size of 64, Adam optimizer [39] with learning rate 0.001, $\beta_1=0$, $\beta_2=0.99$). We train the GAN (batch size of 16, Adam optimizer with learning rate 0.001, $\beta_1=0.9$, $\beta_2=0.99$) until it has seen ten million images, at which point no further improvement for the conditioning loss can be observed.

The evaluation of our model consists of a qualitative and quantitative component: we evaluated how indifferentiable the synthetic MR images are from real MR images and how well the intended contrast settings (through the acquisition parameters TR and TE) are reflected within the synthetic MR images.

## 4.3 Qualitative Evaluation

While different measures have been proposed for the assessment of GANs without reference images (i.e., without corresponding ground truth image pairs), a human observer study remains the gold standard for image quality evaluation [16]. The so-called visual Turing tests are commonly used in computer vision and medical imaging to evaluate how indistinguishable generated images are from real ones [17, 21, 22, 40, 41].

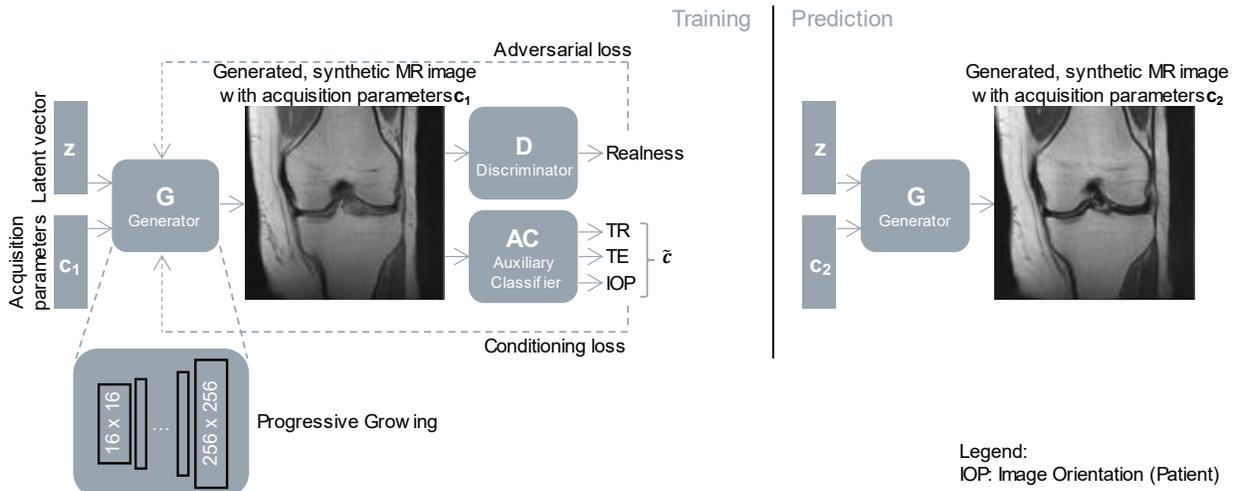

**Fig. 2** Training and inference phase of our GAN. The generator is trained to synthesize MR images for a given latent vector $z$ and a set of acquisition parameters $c_1$, guided by two networks, the discriminator, and the auxiliary classifier. After training, a synthetic MR image can be generated for a given latent vector $z$ with any acquisition parameters $c_2$ (prediction phase). The shown images are generated for a random latent vector with two different sets of acquisition parameters (represented by $c_1$, $c_2$), corresponding to coronal imaging orientation, TR of 3000 ms and a TE of 15 ms and 45 ms for $c_1$ and $c_2$, respectively.

However, when asking medical experts to label randomly displayed images as either real or synthetic in a visual Turing test, the experiment may be subject to bias towards labeling more images as synthetic [22, 21]. Thus, the reported (accuracy) metrics are also biased and less conclusive. Therefore, we adapted the commonly used, biased experiment by using a grid of images ($3 \times 2$ images) with an equal number of real and synthetic images displayed. The evaluator/expert labels each image as real or synthetic (by either clicking the left or the right mouse button) and must mark the same number of images as real and synthetic within each grid. Additionally, for the images marked as synthetic, the expert can explain why the image appears synthetic. This experiment setup removes potential labeling bias by enforcing the true label distribution for the predicted labels.

We asked two experts to label 150 images (75 synthetic, 75 real images; displayed in a random, but the same order for both experts) as either synthetic or real with the experiment mentioned above. The experts have more than 15 years' experience in MRI as an MR technologist (expert 1) and five years' experience as a radiologist (expert 2). No prior information (e.g., examples of synthetic MR images) was provided before the experiment, and only the MR images and the TR and TE values were shown. The TR and TE values and the imaging orientation of the displayed synthetic images matched the values from the real images. No feedback was provided during the experiment on whether the labeling was correct. The confusion matrix of the visual Turing test is presented in Table 2.

**Table 2**
Confusion Matrix – visual turing test

|  |  |  | True label | |
|---|---|---|---|---|
|  |  |  | Real | Synthetic |
| Predicted label | Real | Expert 1 | 53 | 22 |
|  |  | Expert 2 | 36 | 39 |
|  |  | IRA | 29 | 10 |
|  | Synthetic | Expert 1 | 22 | 53 |
|  |  | Expert 2 | 39 | 36 |
|  |  | IRA | 15 | 24 |

IRA: Inter-reader agreement

The experts reached an accuracy score of 71% (expert 1) and 48% (expert 2) on the identification of real and synthetic MR images. The experts were unable to distinguish a significant share of real and synthetic images correctly, which shows the generator's ability to synthesize MR images indistinguishable from real images w.r.t. anatomy and contrast. The low inter-reader agreement for true positives (29/75 images) and true negatives (24/75 images) shows that only the minority of cases could clearly be identified as either real or synthetic.

According to the experts, image quality impediments of synthetic MR images were mainly attributed to overly smooth tissue (muscles, fat tissue, bones) compared to fibrous muscle tissue, granular texture in fat tissue, and fine structures in bones in real MR knee images. Additionally, due to the downsampling of some of the real images to a resolution of 256×256 pixels using bilinear interpolation to obtain identical image resolution, the image quality of the real images was described as inferior in some cases, making it hard for the experts to classify. An example of acquisition parameter interpolation is shown in Fig. 3, and additional examples of varying anatomy demonstrating the variability of the generated samples are shown in Fig. 5. The effect of TE on the tissue contrast can be seen in signal changes of the muscle tissue (Fig. 3). Varying TR results in signal differences in, e.g., the fluid-cartilage contrast [42] and affects mainly contrast on T1-weighted images [43], which are not available within the dataset (see Section 5. Discussion). Therefore, the signal changes through varying TR are less prominent.

## 4.4    Quantitative Evaluation

The performance of a GAN to generate conditional samples depends on proper guidance during training by a well-trained auxiliary classifier. Therefore, we evaluated different conditional GAN architectures w.r.t. the classification and regression performance to determine the acquisition parameters on the test dataset (Table 3). Although the mean squared error was used as regression loss for TR and TE, we report the mean

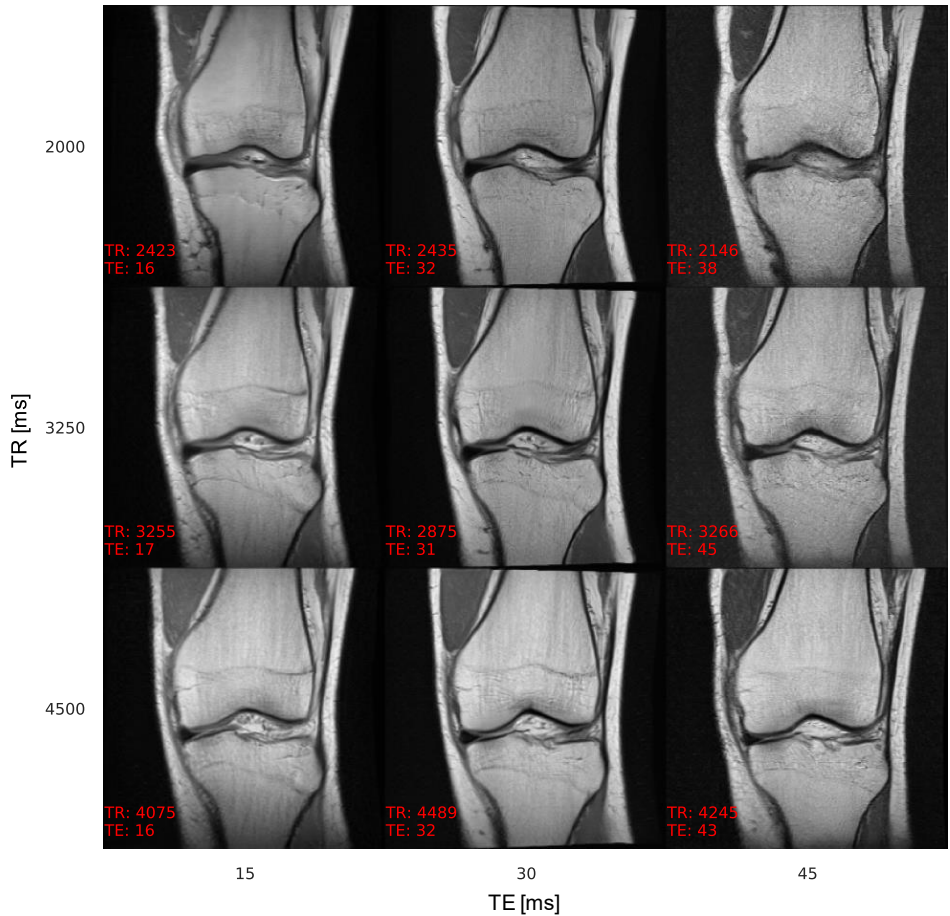

**Fig. 3** Acquisition parameter interpolation of TR and TE for a single latent vector. A single latent vector was reconstructed with different TR and TE values, showing the capability of the generator to synthesize MR images with adaptable image contrast. The axes describe the intended acquisition parameter values and the values at the bottom left of each image the output of the AC. The images are annotated (in red) with acquisition parameter values as determined by the AC, showing a low overall conditioning error. The contrast adapts properly within images along the axes; however, the anatomy also slightly changes, which is a sign of feature entanglement of the latent vector with the conditions.

absolute error (MAE), as the MAE is a more descriptive metric for these targets.

**Table 3**
Auxiliary Classification performance Optimizations

| Model Architecture | Image orientation | TR | TE |
|---|---|---|---|
| | Accuracy [%] | MAE [ms] | MAE [ms] |
| ACGAN | 63.8 | 640.0 | 6.4 |
| Separate AC | <u>100</u> | 225.7 | 1.0 |
| Xception/HP Tuning | <u>100</u> | <u>198.2</u> | <u>0.7</u> |

MAE: mean absolute error.
Statistically significant best values are underlined (P < 0.05; two-sample t-test).

With the original ACGAN architecture, the discriminator is trained to differentiate between real and fake samples and the conditions [6]. However, training the same model on the GAN loss and the conditioning loss leads to training instabilities and, therefore, poor performance in generating realistic images and conditioning the image synthesis. Introducing a separate AC model with the same progressively growing architecture as the discriminator (except for the output layer), which is only trained on the conditions, significantly improves performance. With the additional model selection and hyperparameter tuning, the performance can be further enhanced, leading to an optimized auxiliary classifier performance. Decoupling the development of the auxiliary classifier from the actual generator and discriminator with a separate network breaks down the complexity of the overall architecture. This facilitates additional data augmentation during training and hyperparameter tuning on the AC, significantly improving the GAN's conditioning performance (Table 3).

The adaptive weighting scheme for the conditioning loss based on ControlGAN yields different benefits. It removes the necessity to tune the loss weights for the various conditions manually. Furthermore, it guides the training process of the GAN properly by focusing the training on the conditioning terms that still need improvement (TR, TE) and decreasing the loss weight for conditions that were already learned sufficiently (imaging orientation) (see Fig. 4). It also balances the overall conditioning learning with the GAN loss (WGAN-GP). Therefore, it prevents overfitting the generator on the conditions and produces realistic MR images. The AC's performance on synthetic data (identical size and label distribution as the test set) and the test set are similar (Table 4), which shows that the generator is neither over- nor underfitting on the conditions.

**Table 4**
Auxiliary Classification performance

| Dataset | Image orientation | TR | TE |
|---|---|---|---|
| | Accuracy [%] | MAE [ms] | MAE [ms] |
| Test | 100 | 198.2 | 0.7 |
| Synthetic | 100 | 219.4 | 2.8 |

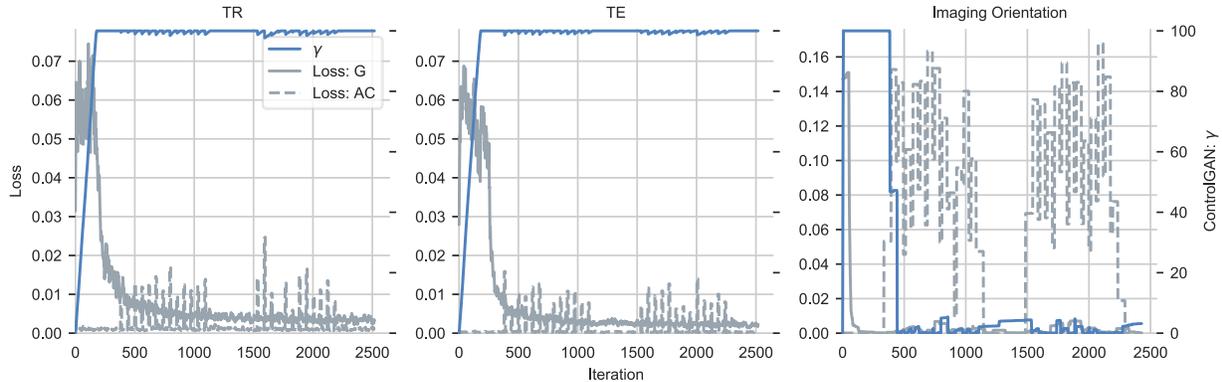

**Fig. 4** Training performance of the GAN for the different conditions including the according adaptive loss weight γ (3). For each acquisition parameters, the adaptive loss weight facilitates the proper conditioning of the generator while preventing overfitting on the conditions. Moreover, it removes the need of manually tuning the loss weight for each acquisition parameter.

## 5. Discussion

Our method generates realistic MR images with high variability in the displayed anatomy that are adjustable in their image contrast through the main contrast acquisition parameter TR and TE. The images are hard to distinguish from real ones for medical experts.

Since no method is currently available to retrieve the TR and TE values from an MR image alone, the quantitative evaluation of the correct contrast is difficult. However, the auxiliary classifier performs well to determine the acquisition parameters on unseen test data and can serve as a reliable method to determine the contrast settings. The GAN's capability to adjust to the contrast settings properly (i.e., TR and TE values) mainly depends on the auxiliary classifier's performance. Consequently, the AC's training has to be improved with additional training data and hyperparameter tuning in future works to enhance the GAN performance further.

The generator's conditioning on the acquisition parameters was limited to values between 1800 ms to 5000 ms for TR and 12 ms to 50 ms for TE due to training data availability. All sequences within the training data were denoted as PD-weighted (by the DICOM series description). A wider range of TR and TE values within additional training data is anticipated to enable T1- and T2-weighted MR image synthesis. However, this dataset has decisive advantages over multiple publicly available datasets [20, 44, 45] despite these limitations. The dataset comprises DICOM files that include MR acquisition parameters and offers a clinically realistic distribution of the acquisition parameters values, as shown in Fig. 1.

Another issue to address in future work is the latent vector's existing feature entanglement, representing the anatomical features, with the conditioning term representing image contrast (and imaging orientation). When adapting the conditions, the image contrast adjusts properly, but the anatomy also slightly changes (see Fig. 3). Feature (dis-)entanglement of generative adversarial networks is a known problem and subject of current research [44].

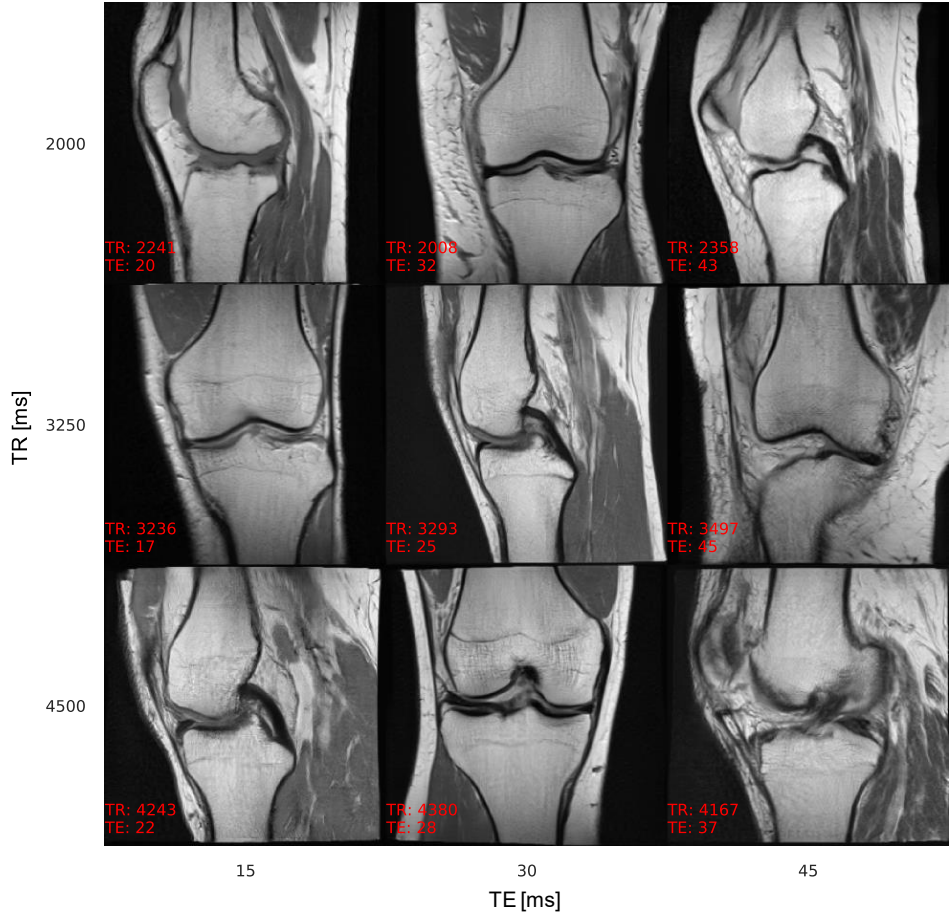

**Fig. 5** Additional examples of synthetic MR images with varying TR and TE to show the variety of the generated image samples. The imaging orientation alternates between sagittal and coronal. The images are annotated (in red) with acquisition parameter values as determined by the AC, showing a low overall conditioning error.

## 5.1 Applications and Future Work

GANs have proven to be a useful data generation and augmentation tool in the medical data domain with the additional benefit of anonymizing patient identifiable information [45]. Therefore, our method can be used as an advanced data augmentation technique for AI training with MR images that can generate MR images with adaptable image contrast. It enables training data generation tailored to an MR application's contrast requirements and is also anticipated to increase AI applications' robustness against contrast changes.

Furthermore, it can be used as a training tool for radiologists and technologists to visualize the influence of the acquisition parameters on the image contrast.

Additionally, it can serve as a tool to support protocol configuration by allowing to preview the yielded contrast for a parameterized sequence. Although medical guidelines exist, MR protocol configuration and sequence parameterization are still a matter of radiologists' and technologists' preferences and vary significantly (see Fig. 1). A tool to visually support sequence parameterization is anticipated to enhance and simplify the protocol configuration workflow. The AC's performance mainly limits the generator's ability to accurately produce an MR image with an arbitrary contrast. The use of simulated MR image data for training of the AC is anticipated to boost the AC's performance further and, therefore, also of the generator. Moreover, the incorporation of additional acquisition parameters (e.g., flip angle, inversion time, scan options) and sequence techniques (e.g., gradient echo) can enhance the model's capability for protocoling support, training data generation, and as a training tool for radiology education. Future work will also include extending our approach to a conditional image-to-image GAN to translate given MR images to new contrast settings.

## 6. Conclusion

We have proposed a generative adversarial model based on the architecture presented in [13] that uses a separate auxiliary classifier and the adaptive conditioning loss from ControlGAN. The network is trained to generate synthetic MR knee images and is conditioned on the MR acquisition parameters TR, TE, and the image orientation. Our approach allows us to generate synthetic MR images that are difficult to distinguish from real images and adapt the MR image contrast based on the input acquisition parameters (TR, TE). Our MR image synthesis approach can adapt to AI-based MR applications' specific requirements by providing fine-tuned, custom MR images with the required contrast. Additionally, it can support radiologists' and technologists' training and be the basis for fine-tuned MR image-to-image contrast translations.

**Acknowledgment**

This work is supported by the Bavarian Academic Forum (BayWISS)—Doctoral Consortium "Health Research", funded by the Bavarian State Ministry of Science and the Arts.